\documentclass[twoside,final]{IEEEtran}
\usepackage{amsmath}
\usepackage{amssymb}
\usepackage[dvips]{graphicx}
\usepackage{cite}
\usepackage{amsthm}

\begin{document}

\newtheorem{proposition}{Proposition}
\newtheorem{corollary}{Corollary}

\title{On the Performance Analysis of RIS-Empowered Communications Over Nakagami-$m$ Fading}
\author{Dimitris Selimis,~\IEEEmembership{Student~Member,~IEEE},
	Kostas~P.~Peppas,~\IEEEmembership{Senior~Member,~IEEE},\\
	George~C.~Alexandropoulos,~\IEEEmembership{Senior~Member,~IEEE}, and Fotis~I.~Lazarakis,~\IEEEmembership{Senior~Member,~IEEE}
    \thanks{D. Selimis and K. Peppas are with the Department of Informatics and Telecommunications, University of Peloponnese, 22131 Tripoli, Greece. D. Selimis is also with Institute of Informatics and Telecommunications, National Center for Scientific Research ``Demokritos,'' 15310 Athens, Greece	(e-mail: \{mst18007, peppas\}@uop.gr).}
    \thanks{G. C. Alexandropoulos is with the Department of Informatics and Telecommunications, National and Kapodistrian University of Athens, Panepistimiopolis Ilissia, 15784 Athens, Greece (e-mail: alexandg@di.uoa.gr).}
       \thanks{F. I. Lazarakis is with the Institute of Informatics and Telecommunications, National Center for Scientific Research ``Demokritos,'' 15310 Athens, Greece (e-mail: flaz@iit.demokritos.gr).}
       \thanks{This work has been partially supported by EU H2020 RISE-6G.}
}

\markboth{IEEE Communications Letters,~Vol.~X,
	No.~XX,~XXXXX}{D. Selimis \MakeLowercase{\textit{et al}.}: Performance Analysis of RIS-Empowered Communications Over Nakagami-\MakeLowercase{$m$} Fading}

\maketitle

\begin{abstract}
In this paper, we study the performance of wireless communications empowered by Reconfigurable Intelligent Surface (RISs) over Nakagami-$m$ fading channels. We consider two phase configuration designs for the RIS, one random and another one based on coherent phase shifting. For both phase configuration cases, we present single-integral expressions for the outage probability and the bit error rate of binary modulation schemes, which can be efficiently evaluated numerically. In addition, we propose accurate closed-form approximations for the ergodic capacity of the considered system. For all considered metrics, we have also derived simple analytical expressions that become tight for large numbers of RIS reflecting elements. Numerically evaluated results compared with Monte Carlo simulations are presented in order to verify the correctness of the proposed analysis and showcase the impact of various system settings.
\end{abstract}


\newcounter{mytempeqncnt}

\section{Introduction}\label{Sec:Intro}
\IEEEPARstart{R}{econfigurable} Intelligent Surfaces (RISs) have recently emerged as a promising candidate technology for next generation wireless systems due to their ability to reconfigure the wireless communication environment in an intelligent manner, thus increasing reception reliability at low cost \cite{J:CHuangIRS,risTUTORIAL2020,Marco2019}. An RIS consists of a large number of ultra-low power consumption elements which are capable of electronically controlling the phase of the incident electromagnetic waves. Recent research results indicate that RISs have great potential due to their promising gains achieved in terms of spectral and energy efficiency without the need of higher hardware complexity and
cost \cite{J:CHuangIRS}. Moreover, RISs have been envisioned as a key enabling
technology to support sixth generation (6G) wireless communications \cite{J:CHuangIRS}. Some typical RIS applications in emerging 6G systems
include multi-user multicast systems, simultaneous wireless information and power transfer (SWIPT),
non-orthogonal multiple access (NOMA) systems, physical-layer security  and cognitive radio (CR) networks, e.g. see \cite{J:CHuangIRS} and references therein.

Because of their promising properties, the performance of RISs over fading channels has been addressed in several recent research works. Representative examples can be found in \cite{J:QinTaoIRS, J:Gunasinghe, J:LiangYangIRS, J:Samuh} and references therein. For example, \cite{J:QinTaoIRS} presented tight bounds and asymptotic results on the performance of RIS systems in the presence of mixed Rayleigh and Ricean fading channels. In \cite{J:Gunasinghe}, bounds on the outage and Ergodic Capacity (EC) performance of RIS-empowered systems operating over fading channels have been presented.
In \cite{J:LiangYangIRS}. accurate approximations to to the end-to-end (e2e) Signal-to-Noise Ratio (SNR) distribution of RIS-assisted systems operatin over Rayleigh fading channels has been presented.
An approximate performance evaluation of RIS-empowered systems over Nakagami-$m$ fading channels is available in \cite{J:Samuh}.

All of the above cited works have considered Optimal Phase Shifting (OPS), i.e., the phase shift of each RIS element is matched with the phases of its incoming and outgoing fading channels. Although such a design yields the optimal performance, it may not be applicable in many cases of practical interest. Specifically, on the one hand, perfect Channel State Information (CSI) at the source is required, resulting in an extensive system overhead. On the other hand, the finite resolution of practical RIS phase shifters poses additional difficulties to the OPS implementation. In \cite{J:Ferreira}, exact expressions and asymptotic results for the BER of large RIS-assisted systems over Nakagami-$m$ fading have been presented, assuming a Von-Mises distribution for the phase error. In \cite{J:Tahir}, a moment-based method was employed to analyze the performance
of RIS-assisted NOMA under Nakagami-$m$ fading with OPS and Random Phase Shifting (RPS). In \cite{J:TaoSZhang}, a random passive beamforming scheme for RIS-assisted multi-user multicast systems have been proposed and approximate performance evaluation expressions have been deduced.
In \cite{C:Psomas}, energy efficient schemes for RIS-aided communications via random phase rotations have been proposed.
 The impact of OPS and RPS on the performance of RIS-based NOMA systems has been addressed in \cite{J:ZhiuoDingNOMAIRS}, where tight approximations for the outage and ergodic capacity performance have been presented. In \cite{J:Houandothers}, large RIS-assisted MIMO systems have been analyzed using a stochastic geometry approach.
Furthermore, because of the inherent difficulty of finding closed-form expressions for the e2e SNR statistics, these works mainly resort to accurate closed-form approximations, tight bounds, or asymptotic analysis. Therefore, as it has also been pointed out in \cite{J:ZhiuoDingNOMAIRS}, finding exact analytical expressions for such statistics is an important research direction. 

Motivated by the above facts, the novel contributions of this letter are summarized as follows.
A comprehensive performance analysis of RIS-empowered systems with RPS and OPS in the presence of Nakagami-$m$ fading is presented. The proposed analytical approach allows for the exact evaluation of important performance metrics, namely the Outage Probability (OP) and the average Bit Error Rate (BER) of binary modulation schemes. Moreover, accurate analytical expressions for EC are presented. Simple and accurate analytical performance evaluation results that become tight for large numbers of RIS unit elements are also deduced. A diversity order analysis for both OPS and RPS schemes is carried out. Our newly derived results can serve as benchmarks to the performance analysis of RIS-empowered systems. All analytical results are substantiated with semi-analytical Monte Carlo simulations.
\footnote{
\emph{Mathematical Notations}:	$\jmath = \sqrt{-1}$, ${\rm Re}\{z\}$ ${\rm Im}\{z\}$ are the real and imaginary parts of the complex number $z$, respectively, $\Gamma \left( \cdot \right) $ is the Gamma function \cite[eq. (8.310/1)]{B:Gradshteyn_00},	$\Gamma \left( \cdot, \cdot \right) $ is the incomplete Gamma function \cite[eq. (8.350/2)]{B:Gradshteyn_00}, $\mathcal{G}$ is the Euler-Mascheroni constant, $J_{a}\left(\cdot\right)$  is the Bessel function of the first kind and order $a$ \cite[eq. (8.402)]{B:Gradshteyn_00}, $\null_pF_q \left( \cdot \right) $ is the generalized hypergeometric function \cite[eq. (9.14/1)]{B:Gradshteyn_00}, and $Q(x)$ is Gauss Q-function. $\mathbb{E} \left\langle \cdot \right\rangle$ denotes expectation, $\mathrm{var}\{\}$ denotes variance and $X\sim \mathcal{N}\left(0,\sigma^2\right)$ represents a zero-mean complex normal Random Variable (RV) with variance $\sigma^2$. $f_X \left( \cdot \right) $ denotes $X$'s Probability Density Function (PDF), $F_X \left( \cdot \right) $ is its Cumulative Distribution Function (CDF),	$\Psi_X \left( \cdot \right) $ is the CHaracteristic Function (CHF) of $X$, and $\Phi_X \left(\omega\right) \triangleq \int_0^\infty J_0(\omega x)f_X(x)\mathrm{d}x $ is the Hankel transform of the PDF of $X$.}
\section{System and Channel Models}\label{Sec:Main}
Let us consider an RIS-assisted dual-hop communication system with one source node
($\mathcal{S}$), an RIS with $N$ reflecting elements, and one destination node ($\mathcal{D}$), as shown in \cite[Fig. 1]{J:HazemIbrahim}. The baseband received signal at $\mathcal{D}$ can be mathematically expressed as \cite[eq. (1)]{J:QinTaoIRS}
\begin{equation}\label{Eq:Smodel}
y = \sqrt{P}x\left(\sum_{n=1}^N h_ng_nr_n + h_d\right)+v,
\end{equation}
where $P$ is the transmit power, $x$ is the unit energy information bearing signal,
$h_n$ and $g_n$ are the flat fading coefficients between $\mathcal{S}$ and the $n$-th RIS reflecting element and between the $n$-th RIS element and $\mathcal{D}$, respectively, and $h_d$ is the flat fading coefficient between $\mathcal{S}$ and $\mathcal{D}$. In addition, $r_n = \exp(\jmath \theta_n)$ is the $n$-th RIS element response with $\theta_n \in [0, 2\pi]$ being the phase shift, and $v$ is the additive Gaussian noise having zero mean and variance $\sigma_n^2$. In \eqref{Eq:Smodel}, $h_n = |h_n|\exp(\jmath \theta_{h_n})$, $g_n = |g_n|\exp(\jmath \theta_{g_n})$ and $h_d = |h_d|\exp(\jmath \theta_{h_d})$ are assumed to be independent RVs whose envelopes $|h_n|$, $|g_n|$, and $|h_d|$ follow a Nakagami-$m$ distribution with parameters $(m_{h_n}, \Omega_{h_n})$,  $(m_{g_n}, \Omega_{g_n})$, and $(m_{h_d}, \Omega_{h_d})$, respectively. The parameters $\Omega_{\ell}$ for $\ell \in \{h_n, g_n, h_d\}$,
are given as $\Omega_\ell = \zeta r_\ell^{-\alpha}$ where $\zeta = \left(\frac{\lambda}{ 4\pi}\right)^2$ is the near-field pathloss factor at a reference distance of 1 m, with $\lambda$ being the carrier wavelength, and $\alpha \geq 2$ is the pathloss exponent. The distances $r_\ell$, $\forall$$\ell$, satisfy the cosine law, namely, $r_{h_d}^2 = r_{h_n}^2+ r_{g_n}^2-2r_{h_n}r_{g_n}\cos(\psi_n)$, where $\psi_n$ is the angle between
the $\mathcal{S}$-RIS and RIS-$\mathcal{D}$ links. Finally, the phases $\theta_{h_n}$, $\theta_{g_n}$, and $\theta_{h_d}$ are assumed to be independent to each other, as well as independent to the envelopes $|h_n|$, $|g_n|$ and $|h_d|$ \cite{J:YacoubPhaseEnv}.

When CSI is not available at $\mathcal{S}$, i.e., the phases
$\theta_{h_n}$ and $\theta_{g_n}$ are not known to the RIS,
RPS can be employed. This design reduces the system overhead needed for acquiring CSI \cite{J:ZhiuoDingNOMAIRS}.
The resulting e2e SNR is given as
\begin{equation}\label{Eq:Random}
\gamma_R = \rho\left|\sum_{n=1}^N \left|h_n \right| \left|g_n\right|\exp(\jmath \phi_n) + \left|h_d\right|\exp(\jmath \phi_{d})\right|^2,
\end{equation}
where $\rho = \frac{P}{\sigma_n^2}$, $\phi_n = \theta_{h_n}+\theta_{g_n}-\theta_n$, and $\phi_d = \theta_{g_n}-\theta_d$ with $\theta_n$ and $\theta_d$ being randomly chosen phases.

On the other hand, assuming perfect CSI, the RIS can apply the OPS scheme. In this case, each $\theta_n$ is matched with $\theta_{h_n}$ and $\theta_{g_n}$, whereas $\theta_d$ is matched with $\theta_{h_d}$ yielding the following e2e SNR expression:
\begin{equation}\label{Eq:Coherent}
\gamma_C = \rho\left(\sum_{n=1}^N \left|h_n \right| \left|g_n\right|+\left|h_d\right|\right)^2.
\end{equation}
In the following analysis, the statistics of $\gamma_R$ and $\gamma_C$ will be analytically characterized.

\section{End-to-End SNR Statistics}\label{Sec:SNRStatistics}
When RPS is employed, it can be observed that the evaluation of the statistics of $\gamma_R$ requires the evaluation of the statistics of the sum of $N$ complex RVs with amplitudes $X_n =\left|h_n \right|\left|g_n\right|$ and phases $\phi_n$.
In the problem under consideration,
the amplitudes $X_n$ follow a double Nakagami-$m$ distribution with PDF given by \cite[eq. (8)]{J:Ferreira}.
The $n$-th moments of $X_n$ are given as \cite[eq. (9)]{J:HazemIbrahim}
\begin{equation}\label{Eq:moms2}
\mu_{X_n}(k) = \Lambda_n^{k/2}\frac{\Gamma\left(m_{h_n}+\frac{k}{2}\right)\Gamma\left(m_{g_n}+\frac{k}{2}\right)}{\Gamma(m_{h_n})\Gamma(m_{g_n})},
\end{equation}
where $\Lambda_n = \frac{\Omega_{h_n}\Omega_{g_n}}{m_{h_n} m_{g_n}}$.
According to \cite{J:YacoubPhaseEnv}, the phases $\theta_{h_n}$, $\theta_{g_n}$, and $\theta_{h_d}$ are independent to $|h_n|$, $|g_n|$ and $|h_d|$, respectively, and their PDFs are given by \cite[eq. 3]{J:YacoubPhaseEnv}.
Moreover, it is assumed that $\theta_{n}$ is uniformly distributed in $[0, 2\pi]$.

An analytical expression for the PDF of $\phi_n$ is very difficult to be derived in closed form, mainly due to the difficulty in finding the PDF of the sum of one uniform RV and two RVs distributed according to \cite[eq. 3]{J:YacoubPhaseEnv}. In order to address this difficult problem, it is hereinafter assumed that $\phi_n$ is uniformly distributed in $[0, 2\pi]$. As it will become evident in Section \ref{Sec:Numerical}, this approximation yields very accurate results that are almost indistinguishable to the exact distribution.
Using \cite[eq. (10)]{J:AbdiRandomVectors} and employing a transformation of RVs, the PDF of $\gamma_R$ can be deduced as
\begin{equation}\label{eq:1}
f_{\gamma_R}(\gamma) = \frac{1}{2\rho}\int_0^{\infty}tJ_0\left(t\sqrt{{\gamma}/{\rho}}\right)\mathcal{H}(t)\mathrm{d}t,
\end{equation}
where $\mathcal{H}(t) = \Phi_{|h_d|}(t)\prod_{n=1}^N\Phi_{X_n}(t)$. By employing
\cite[eqs. (3.915.2) and (6.631.1)]{B:Gradshteyn_00} and \cite[eq. (6.576/3)]{B:Gradshteyn_00}, the Hankel transforms $\Phi_{|h_d|}(t)$ and $\Phi_{X_n}(t)$ can be obtained as
\begin{subequations}\label{eq:2}
\begin{equation}
\Phi_{|h_d|}(t) = \null_1F_1\left(m_{h_d};1; -0.25  \Lambda_d t\right),
\end{equation}
\begin{equation}
\Phi_{X_n}(t) = \null_2F_1\left(m_{h_n},m_{g_n};1; -0.25 \Lambda_n t^2\right),
\end{equation}
\end{subequations}
where $\Lambda_d = \frac{\Omega_{h_d}}{m_{h_d}}$.
Using \cite[eq. (5.56/2)]{B:Gradshteyn_00}, \eqref{eq:1}, and \eqref{eq:2}, the CDF of $\gamma_R$ is given after some algebraic operations as
\begin{equation}\label{Eq:CDFrnd}
F_{\gamma_R}(\gamma) = \sqrt{{\gamma}/{\rho}}\int_{0}^{\infty} J_1\left(\sqrt{{\gamma}/{\rho}}t\right)\mathcal{H}(t)\mathrm{d}t.
\end{equation}
Finally, the $k$-th moment of $\gamma_R$ can be obtained in the following closed form \cite[eq. (23)]{J:AbdiRandomVectors}:
\begin{equation}\label{eq:momentsR}
\mu_{\gamma_R}(k) = \left.(-4)^{k} \frac{(k!)^2}{(2k)!}\frac{\partial^{2k}}{\partial t^{2k}} \mathcal{H}(t)\right|_{t=0}.
\end{equation}

We hereinafter derive closed-form statistics of the RV $Y = \sqrt{\rho}\sum_{n=1}^NX_n \exp{(\jmath\phi_n) }$ for large values of $N$. In what follows, it is assumed that $m_{h_n} = m_h$ and $m_{g_n} = m_g$, $\forall n$. Moreover, we make use of the distances approximations $r_{h_n} \approx r_h$ and $r_{g_n} \approx r_g$, which imply that $\Omega_{h_n} \approx \Omega_{h}$ and $\Omega_{g_n} \approx \Omega_{g}$. Let also $Y^R_n = \mathrm{Re}\{X_n \exp{(\jmath\phi_n) }\}$ and
$Y_n^I = \mathrm{Im}\{X_n \exp{(\jmath\phi_n) }\}$. By observing that $Y^R_n$ is independent to $Y^R_m$ for $n \neq m$ and employing the Central Limit Theorem (CLT), $\mathrm{Re}\{Y\} \sim \mathcal{N}\left(0,\sigma_1^2\right)$ with $\sigma_1^2 = N\rho\Omega_g\Omega_h/2$. It follows that $\mathrm{Im}\{Y\}$ follows the same distribution as $\mathrm{Re}\{Y\}$. Moreover, following a similar line of arguments as in \cite{J:ZhiuoDingNOMAIRS}, it can be deduced that $\mathrm{Re}\{Y\}$ and $\mathrm{Im}\{Y\}$ are independent. Thus, the RV $|Y|^2$ follows an exponential distribution with PDF $f_{|Y|^2}(x) = 1/(2\sigma_1^2)\exp(-x/(2\sigma_1^2))$ and CDF $F_{|Y|^2}(x) = 1-\exp(-x/(2\sigma_1^2))$. Hence, the CHF of $|Y|^2$ can be deduced as $\Psi_{|Y|^2}(t) = (1-\jmath 2\sigma_1^2 t)^{-1}$.
Moreover, it can be deduced that the mean of $|Y|^2$ equals to $N \Omega_g \Omega_h \rho$. Thus, the e2e SNR under RPS increases linearly with $N$.

For OPS, a closed-form expression for the statistics of $\gamma_R$ is very difficult - if not impossible - to be derived. Nevertheless, by employing a CHF-based approach, analytical expressions for the statistics of the e2e SNR $\gamma_R$ can be readily obtained.
Specifically, by employing the inversion theorem \cite{J:GilPelaez}, an integral expression for the CDF of $\gamma_R$ is obtained as
\begin{equation}\label{Eq:CDFcoherent}
F_{\gamma_C}(\gamma) = \frac{1}{2}-\frac{1}{\pi}\int_0^{\infty}t^{-1}{\rm Im}\left\{\exp\left(-\jmath t \sqrt{\frac{\gamma}{\rho}}\right) \Psi_{\gamma_C}(t) \right\}\mathrm{d}t.
\end{equation}
Assuming that $X_n$'s are statistically independent, results in $\Psi_{\gamma_C}(t) = \Psi_{|h_d|}(t)\prod_{n=1}^N \Psi_{X_n}(t)$, where
$\Psi_{|h_d|}(t)$ and $\Psi_{X_n}(t)$ are given respectively by \cite{J:Annamalai_00} and \cite{J:Peppas3}
\begin{align}
&\Psi_{|h_d|}(t) = \null_1F_1\left(m_{h_d}; 0.5; -0.25 \Lambda_d t^2\right) \nonumber
\\ & +\frac{\jmath t\sqrt{\Lambda_d}\Gamma(m_{h_d} + 0.5)}{\Gamma(m_{h_d})}
  \null_1F_1(m_{h_d} + 0.5; 1.5; -0.25\Lambda_d t^2), \nonumber
\end{align}
\begin{align}
&\Psi_{X_n}(t)  =
\frac{(16/\Lambda_n)^{m_{h_n}}\Gamma\left(m_{h_n}+0.5\right)\Gamma\left(m_{g_n}+0.5\right)}{\sqrt{\pi}\left(-\jmath t+2/\sqrt{\Lambda_n}\right)^{2 m_{h_n}}\Gamma\left(m_{h_n}+m_{g_n}+0.5\right)} \nonumber \\
&  \times \left.\null_2F_1\left(2m_{h_n}, m_{h_n}-m_{g_n}+0.5; m_{h_n}+m_{g_n}+0.5; \right.Z(t) \right),\nonumber
\end{align}
with $Z(t) = \frac{-\jmath t-2/\sqrt{\Lambda_n}}{-\jmath t+2/\sqrt{\Lambda_n}}$.
Finally, the $k$-th moment of ${\gamma_C}$ can be easily expressed in terms of the moments of $X_n$ by employing the multinomial theorem, yielding
\begin{align}\label{Eq:momsZ}
&\mu_{\gamma_C}(k) = \rho^k\sum_{j_1=0}^{2k}\sum_{j_2=0}^{j_1}\cdots\sum_{j_{N-1}=0}^{j_{N-2}}
\binom{2k}{j_1}\binom{j_1}{j_2}\cdots\binom{j_{N-2}}{j_{N-1}} \nonumber \\
&\times \mu_{X_1}(2k-j_1)\mu_{X_2}(j_1-j_2)\cdots\mu_{X_N}(j_{N-1}).
\end{align}
Next, we derive closed-form statistics of the RV $\tilde{Z} = \sum_{n=1}^NX_n$ for large values of $N$.
Under the same assumptions as in the RPS case and by employing CLT, it yields $\tilde{Z}\sim \mathcal{N}(N \mathbb{E}\langle\tilde{Z}\rangle, N \mathrm{var}\{\tilde{Z}\})$, where $E\langle\tilde{Z}\rangle$ and $\mathrm{var}\{\tilde{Z}\}$ can be directly evaluated from \eqref{Eq:moms2}. It, hence, follows that $Z = \rho\tilde{Z}^2$ follows a non-central chi-square distribution with the following PDF:
 \begin{align}
 f_Z(x) = s\exp[-0.5(\xi -s x)]\cosh(\sqrt{\lambda s x})/\sqrt{2\pi s x}
 \end{align}
  where
$\xi = N \mathbb{ E}^2\langle\tilde{Z}\rangle/\mathrm{var}\{\tilde{Z}\}$
and $s^{-1} = \rho N \mathrm{var}\{\tilde{Z}\}$.
The CDF of $Z$ can be deduced after performing some straightforward manipulations as
\begin{align}
F_{Z}(x) = Q(\sqrt{\xi}- \sqrt{xs}) - Q(\sqrt{\xi} + \sqrt{x s}).
\end{align}
Then, the CHF of $Z$ can be derived as $\Psi_{Z}(t) = \exp[- \xi t/(\jmath s+2 t)] (1-2\jmath t/s)^{-1/2}$, and the mean of $Z^2$ is obtained in closed-form as $N \rho (\mathbb{E}^2\langle \tilde{Z}\rangle N + \mathrm{var}\{\tilde{Z}\})$. We thus conclude that the e2e SNR under OPS increases quadratically with $N$. Finally, it is worth pointing out that, although this asymptotic behaviour for large-$N$ has already been reported in previous research works, assuming Rayleigh or Nakagami-$m$ channels, e.g. see \cite{J:Gunasinghe}, it is evident that such results hold even for arbitrary fading distributions, provided that the CHF or the Hankel transform of the cascaded fading distribution is readily available. Such transforms, enable the straightforward computation of the $n$-th moment of the e2e SNR (see for example \eqref{eq:momentsR}), thus providing a unified framework for the performance analysis of RIS-empowered systems for both small and large values of $N$.

\section{Performance Analysis}\label{Sec:Performance}
\subsubsection{Outage Probability (OP)}
The OP is defined as the probability that the instantaneous SNR does not fall below an outage threshold, $\gamma_{\rm{th}}$. The OP can be expressed in terms of the CDF of the instantaneous SNR, $\gamma$, as $P_{\rm out} = F_{\gamma}(\gamma_{\rm{th}})$. Thus, $P_{\rm out}$ can be readily evaluated for both RPS and OPS designs by employing \eqref{Eq:CDFrnd} and \eqref{Eq:CDFcoherent}, respectively. It is pointed out that although the integrand in \eqref{Eq:CDFrnd} is oscillatory due to the presence of the Bessel function, the evaluation of OP can be performed in a reliable and computationally efficient manner using standard built-in routines available in popular mathematical packages, e.g. Mathematica. For large values of $N$ and assuming absence of the direct link, OP can be efficiently calculated using the analysis of the previous section.
\subsubsection{Average Bit Error Rate (BER)}
The average BER of binary modulations schemes can be expressed in terms of the CDF of the average SNR as follows \cite{B:Alouini}:
\begin{equation}\label{Eq:BERCDF}
P_{be} = \frac{q^p}{2\Gamma(p)}\int_0^{\infty}\gamma^{p-1}\exp(-q\gamma)F_{\gamma}(\gamma)\mathrm{d}\gamma,
\end{equation}
where $(p,q) = (1/2,1)$ for binary phase shift keying (BPSK), $(p,q) = (1/2,1/2)$ for binary frequency shift keying (BFSK), and $(p,q) = (1,1)$ for binary differential phase shift keying (BDPSK).
For RPS, by substituting \eqref{Eq:CDFrnd} into \eqref{Eq:BERCDF} and changing the order of the integration, one obtains
\begin{align}
P_{be, R} =& \frac{q^p}{2\Gamma(p)\sqrt{\rho}}\int_0^{\infty}
\left[\int_{0}^{\infty} J_1\left(\sqrt{\frac{\gamma}{\rho}}t\right)
\gamma^{p-0.5}e^{-q\gamma}\mathrm{d}\gamma
\right] \nonumber \\
& \times \mathcal{H}(t)\mathrm{d}t.
\end{align}
By employing \cite[eq. (6.621/1)]{B:Gradshteyn_00} and performing some straightforward manipulations, a single-integral expression for $P_b$ can be deduced as
\begin{align}
P_{be, R} & = \frac{p q^{1-p}}{4{\rho}}\int_0^{\infty}
t \null_1F_1\left(1+p;2;-\frac{t^2}{4q\rho}\right)
 \mathcal{H}(t)\mathrm{d}t.
\end{align}
Finally, for large values of $N$ and assuming absence of the direct link, by substituting the approximate CDF of the e2e SNR into \eqref{Eq:BERCDF} and performing some straightforward manipulations, a simple closed-form expression for BER can be deduced as $P_{be, R} = 0.5 - 0.5q^p(q + 0.5\sigma_1^{-2})^{-p}$. When the direct link is present, BER can be evaluated using the proposed CHF-based approach and the analysis presented in the previous section.

For OPS and assuming coherent binary modulation schemes, $P_b$ can be obtained using \cite[eq. (19)]{J:Annamalai_00} as:
\begin{align}
P_{be,C}  = \frac{1}{2}-\frac{1}{\pi}\int_{0}^{\infty}t^{-1}\exp(-t^2/2) \mathrm{Im}\{\Psi_{\gamma_{R}}(\sqrt{2\,a\,\rho}t)\}\mathrm{d}t, \nonumber
\end{align}
where $a = 1$ for BPSK and $a = 1/2$ for BFSK. For large-$N$, BER can be obtained using the proposed CHF-based approach and the results presented in the previous section.

\subsubsection{Diversity Order Analysis}
Hereafter, we assume that the direct link is absent for all considered phase combining designs.
For RPS and for high values of $\rho$, by observing that $\null_1F_1\left(1+p;2;-\frac{t^2}{4q\rho}\right) \rightarrow 1$, an asymptotic expression for $P_{be, R}$ can be deduced as
\begin{align}
P_{be, R} \approx \frac{p q^{1-p}}{4{\rho}}\int_0^{\infty}
t \mathcal{H}(t)\mathrm{d}t.
\end{align}
The latter expression reveals that the diversity order of the considered scheme equals to one.
For OPS, the diversity order can be readily deduced as the diversity order of an equal gain combining system operating in the presence of double Nakagami-$m$ fading. Following \cite{C:CindyZhu}, its diversity order equals to $\min \left\{\sum_{n=1}^N m_{h_n},\sum_{n=1}^N m_{g_n}\right\}$.

\subsubsection{Ergodic Capacity (EC)}
The average (ergodic) capacity of the considered system is given by
$\mathbb{E}\langle C \rangle = \frac{1}{\ln 2}\mathbb{E}\langle \ln(1+\gamma)\rangle.$
In what follows, an accurate approximation for the EC evaluation will be presented, which is valid for both RPS and OPS designs. In particular, using a Taylor series approximation around
the first moment of $\gamma$, the EC can be approximated as
\begin{equation}
\mathbb{E}\langle C \rangle \approx \frac{1}{\ln 2}\left[ \ln(1 + \mu(1)) - \frac{ \mu(2) - \mu(1)^2}{2(1 + \mu(2))^2}\right].
\end{equation}
As it will become evident, this approximation is very tight for a wide range of system parameters.
Moreover, for RPS and large values of $N$, by evaluating the expectation and using \cite[eq. (4.337/2)]{B:Gradshteyn_00}, a closed-form expression for the EC can be deduced as
$\mathbb{E}\langle C \rangle \approx \frac{1}{\ln 2}\exp(1/2\sigma_1^2)\Gamma(0, 1/2\sigma_1^2)$. When $\rho\rightarrow \infty$, EC can be further approximated using \cite[eq. (8.212/1)]{B:Gradshteyn_00} as $\mathbb{E}\langle C \rangle \approx \frac{1}{\ln 2}\left(\ln(\rho)+\ln(N\Omega_g\Omega_h)-\mathcal{G}\right)$, which reveals that the high-SNR EC slope equals to one, and the high-SNR power offset scales logarithmically with $N$.

\section{Numerical and Simulation Results}\label{Sec:Numerical}
In this section, numerically evaluated results accompanied with semi-analytical Monte-Carlo simulations
are presented to evaluate the performance of the two considered
transmission schemes. Unless otherwise stated, carrier frequency equals to 2.45 GHz, $\psi = 86^0$ when a direct link is present, $\alpha = 2.5$ and noise power equals to -85 dBm.

\begin{figure}[h]
\centering\vspace{0.4cm}
\includegraphics[keepaspectratio,width=2.4in]{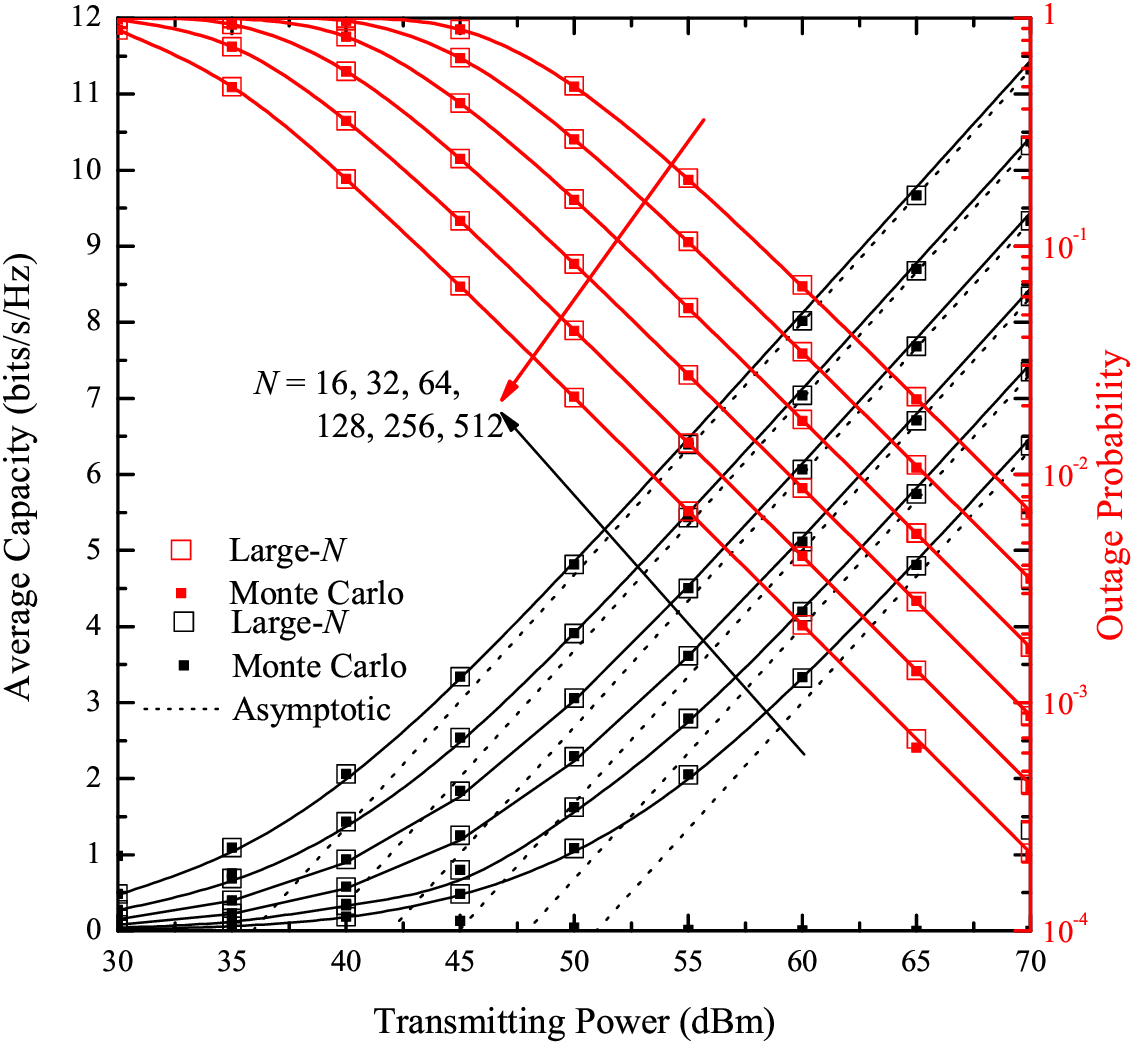}
\caption{OP and EC of RIS-empowered systems with RPS operating in a LOS environment with $K = 10$, as a function of $P$ for various values of $N$. No direct link is present.}\vspace{-0.4cm} \label{Fig:OPECrandom}
\end{figure}

Figure~\ref{Fig:OPECrandom} depicts the OP and EC of the proposed scheme with RPS and without the presence of the direct link as a function of $P$ for various values of $N$. It is also assumed that $r_{h} = r_{g} = 20 \mathrm{m}$ and the involved wireless channels are Line-Of-Sight (LOS) ones. In such environments, the Nakagami-$m$ distribution can be used to approximate the Ricean distribution. The corresponding $m$ parameters can be obtained as
$m = \left(1-\left(\frac{K}{1+K}\right)^2\right)^{-1}$, where $K$ is the Ricean factor \cite{B:Alouini}. Hereafter, it is assumed that $K = 10$.
 In the same figure, equivalent results obtained using semi-analytical Monte-Carlo simulations are also presented, using both the exact distribution of $\phi_n$ and the uniform phase approximation.
As it can be observed, the uniform phase approximation is highly accurate, yielding results that are practically identical to the ones obtained using the exact phase distribution. According to our observation, this is because of the fact that the PDF of the sum of the random phases well approximates the uniform distribution. Moreover, the OP expression evaluations perfectly agree with the simulated OP, whereas the proposed EC approximation is very tight within the entire SNR region. In addition, the proposed large-$N$ and high-SNR analysis yields very accurate results for all considered test cases.

\begin{figure}[h]
\centering\vspace{0.4cm}
\includegraphics[keepaspectratio,width=2.4in]{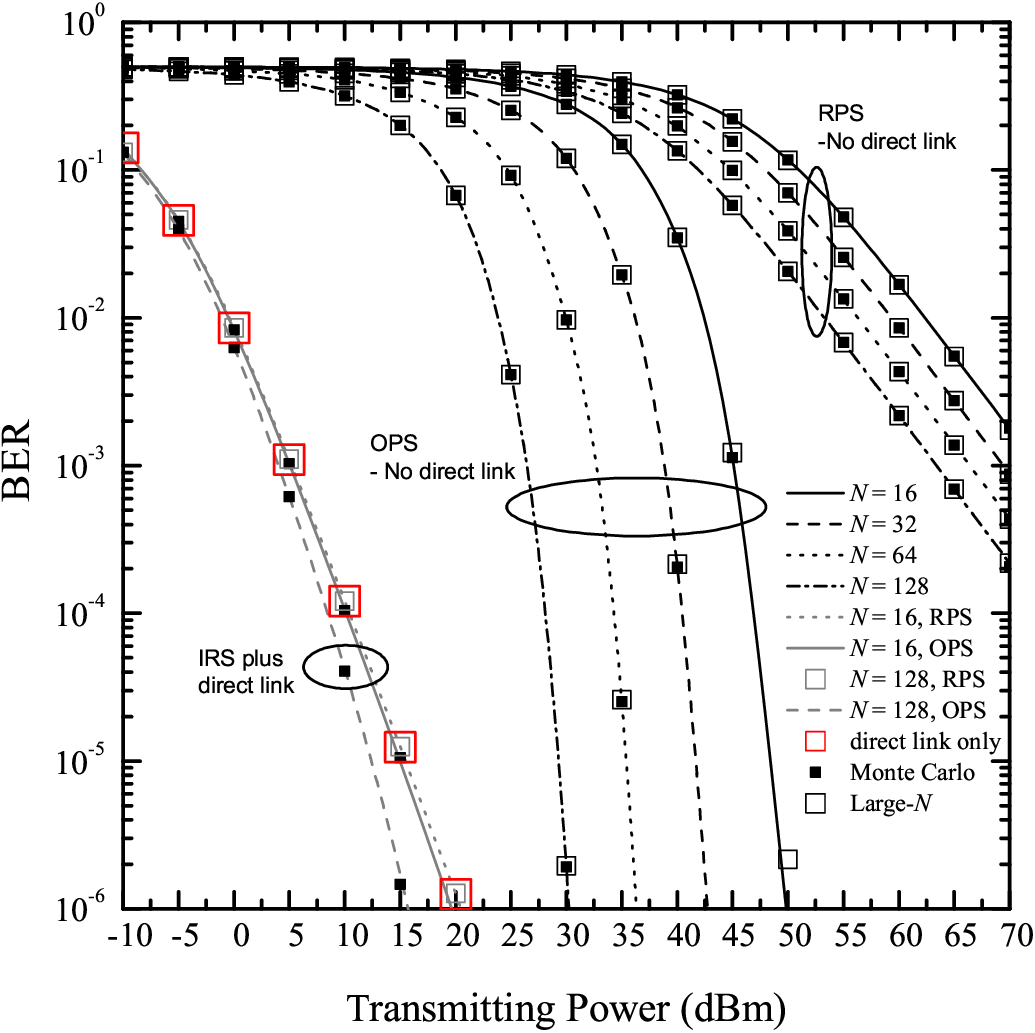}
\caption{Average BER of RIS-empowered systems with RPS and OPS, employing BPSK modulation and operating over Nakagami-$m$ fading, as a function of $P$ for various values $N$, as well as $m_{h_n} = 1.5$ and $m_{g_n} = 2.5$ $\forall n \in \{1,2, \ldots, N\}$. }\vspace{-0.4cm} \label{Fig:BER}
\end{figure}

\begin{figure}[h]
\centering\vspace{0.4cm}
\includegraphics[keepaspectratio,width=2.4in]{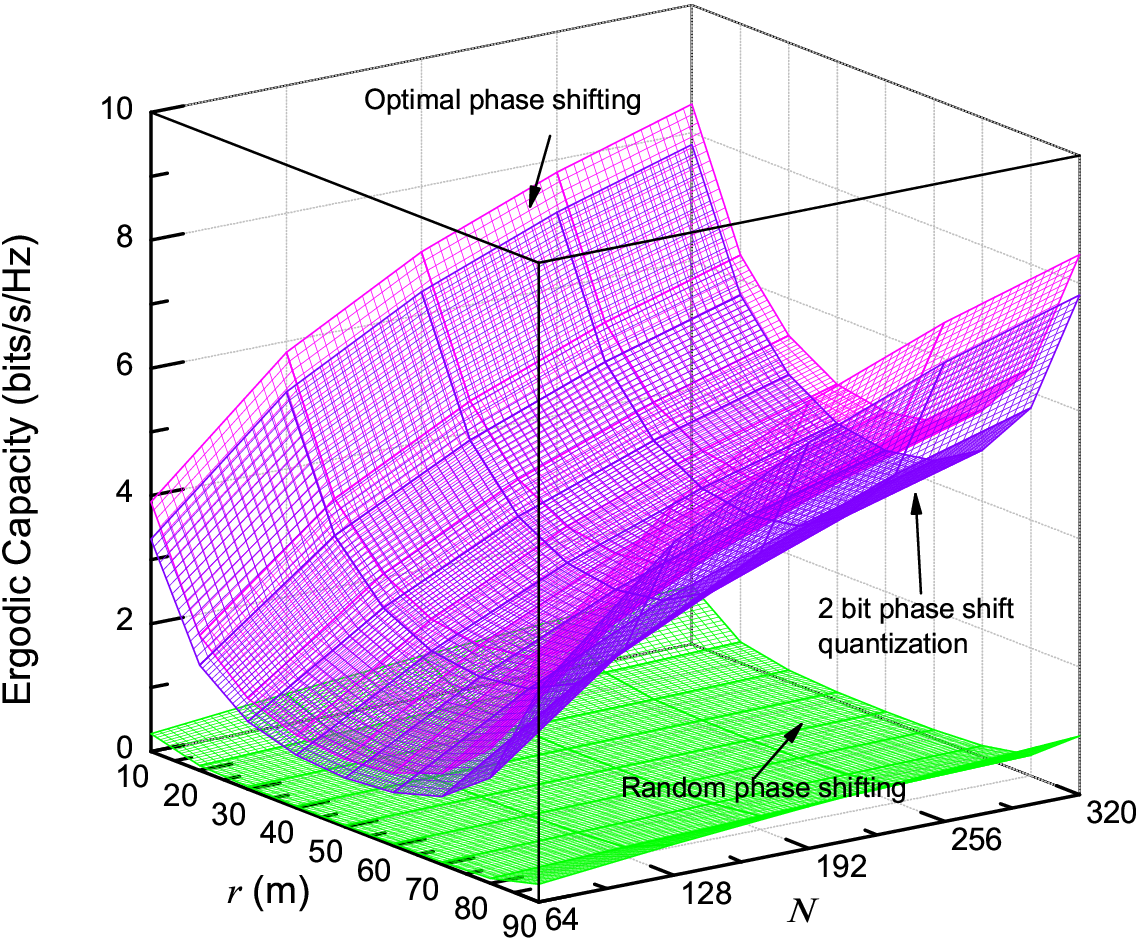}
\caption{EC of RIS-empowered systems as a function of $N$ and the distance $r_h$ of the $\mathcal{S}\rightarrow \mathrm{RIS}$ link.}\vspace{-0.4cm} \label{Fig:CapVsNvsr}
\end{figure}

In Fig.~\ref{Fig:BER}, the average BER of BPSK modulation assuming RPS and OPS, and the presence or absence of the direct link are illustrated.
It is assumed that $r_{h} = r_{g} = 20 \mathrm{m}$ (resulting in $r_{h_d}$ of 27.3 $\mathrm{m}$),
$m_{h_n} = 1.5$, and $m_{g_n} = 2.5$ $\forall n \in \{1,2, \ldots, N\}$. As it can be observed, the OPS scheme significantly outperforms the RPS scheme when no direct link is present, even for small number of RIS unit elements $N$. It is evident that the diversity order of the RPS scheme is independent to the number $N$, as well as to the fading parameters. It is also evident that the presence of the direct link significantly improves system performance. When the RPS scheme is employed, however, performance enhancements are small as $N$ increases. It can also be seen that the overall system performance is comparable to that when only the direct link is present. This is due to the fact that the diversity order of the RPS scheme is independent to the number of the RIS elements, $N$, and primarily affected by the fading parameter of the direct link. When OPS is employed, however, a better performance can be achieved, nevertheless performance gains are smaller compared to the case when direct link is absent.
Finally, it is noted that, in practice, employing continuous phase shifting schemes at the RIS elements
may be prohibitively complicated, due to various hardware limitations.
To this end, practical RIS configurations may employ quantized phase shifts.
In this case, the e2e SNR is of the form of \eqref{Eq:Random}, where $\phi_n$ is the error between the un-quantized and
quantized phase shifts \cite{J:Gunasinghe}.
Assuming a large number of quantization levels, $\phi_n$ follows a uniform distribution in $[-\pi/2^b, \pi/2^b)$,
where $b$ is the number of quantization bits \cite{J:Gunasinghe}.
The EC of RPS, 2-bit quantized phase shifting and OPS designs, as a function of the distance $r_h$ of the $\mathcal{S}\rightarrow \mathrm{RIS}$ link and $N$, with $K = 10$, is demonstrated in Fig.~\ref{Fig:CapVsNvsr}.
The distance between $\mathcal{S}$ and $\mathcal{D}$ equals to 100 m and no direct link is considered. 
It is evident that in all designs, a minimum value of EC is observed when $r_h^\star \approx 50$ m. Also, the quantized scheme performs much better that the RPS design for all values of $N$. Moreover, for $N = 320$ the EC loss performance is approximately 0.6 b/s/Hz as compared to the OPS design.

\section{Conclusions}\label{Sec:Conclusions}
In this paper, the performance of RIS-empowered wireless communication systems with two phase shifting designs over Nakagami-$m$ fading has been studied. Exact analytical expressions and accurate closed-form approximations for important performance metrics were presented.
It has been observed that, in contrary to the optimal design, the diversity order of the random scheme is always one, i.e., independent of the number of RIS unit elements and the fading parameters. 
Numerically evaluated and simulated results were presented highlighting the different tradeoffs between performance and complexity.

\bibliographystyle{IEEEtran}

\end{document}